\documentclass[journal,onecolumn,12pt,twoside]{IEEEtran}

\usepackage[T1]{fontenc}
\usepackage{graphicx}
\usepackage{cite}
\usepackage{caption}
\usepackage{subcaption}

\usepackage{overpic}
\usepackage{amssymb}
\usepackage{amsmath}
\usepackage{amsthm}
\usepackage{array}
\usepackage{color}
\usepackage{url}

\IEEEoverridecommandlockouts

\theoremstyle{plain}

\newcommand{\argmin}[1]{{\underset{{#1}}{\mathrm{arg\,min}}}}
\newcommand{\vect}[1]{\boldsymbol{#1}}

\def\train{\mathrm{train}}

\begin{document}

\title{Two Applications of Deep Learning in the Physical Layer of Communication Systems}

\author{Emil Bj\"ornson and Pontus Giselsson\thanks{E.~Bj\"ornson is with Link\"oping University, Sweden. P.~Giselsson is with Lund University, Sweden. This work was partially supported by the Wallenberg AI, Autonomous Systems and Software Program (WASP) funded by the Knut and Alice Wallenberg Foundation.}}

\maketitle

\IEEEpeerreviewmaketitle

\section*{}
\vskip-2cm

Deep learning has proved itself to be a powerful tool to develop data-driven signal processing algorithms for challenging engineering problems. By learning the key features and characteristics of the input signals, instead of requiring a human to first identify and model them, learned algorithms can beat many man-made algorithms. In particular, deep neural networks are capable of learning the complicated features in nature-made signals, such as photos and audio recordings, and use them for classification and decision making.

The situation is rather different in communication systems, where the information signals are man-made,  the propagation channels are relatively easy to model, and we know how to operate close to the Shannon capacity limits. Does this mean that there is no role for deep learning in the development of 
future communication systems?

\section{Relevance}

The answer to the question above is ``no'' but for the aforementioned reasons, we need to be careful not to reinvent the wheel. We must identify the right problems to tackle with deep learning and, even then, not start from a blank sheet of paper. There are many signal processing problems in the physical layer of communication systems that we already know how to solve optimally, for example, using well-established estimation, detection, and optimization theory. Nonetheless, there are also important practical problems where we lack acceptable solutions, for example, due to a lack of appropriate models or algorithms. In this lecture note, we first introduce the key properties of artificial neural networks and deep learning.  The focus is not on technicalities around the training process or choice of network structure, but on what we can practically achieve, assuming the training is carried out successfully. We will then describe three application categories in communication engineering, whereof one exposes some fundamental weaknesses of deep learning and two illustrate important advances that can be made by utilizing deep learning.

\section{Prerequisites}

This lecture note requires basic knowledge of linear algebra, digital communications, and probability.

\begin{figure} 
        \centering
        \begin{subfigure}[t]{\columnwidth} \centering 
	\begin{overpic}[width=0.6\columnwidth,tics=10]{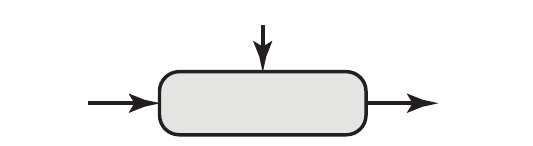}
		\put(84,7){$\vect{y}$}
		\put(43,7){$\hat{f}(\vect{x}_0;\vect{\theta})$}
		\put(10,7){$\vect{x}_0$}
		\put(48.5,24.5){$\vect{\theta}$}
\end{overpic} 
                \caption{An arbitrary gray box taking $\vect{x}_0$ as input and giving $\vect{y}$ as output.} 
        \end{subfigure} 
        \begin{subfigure}[t]{\columnwidth} \centering  \vspace{+2mm}
	\begin{overpic}[width=0.6\columnwidth,tics=10]{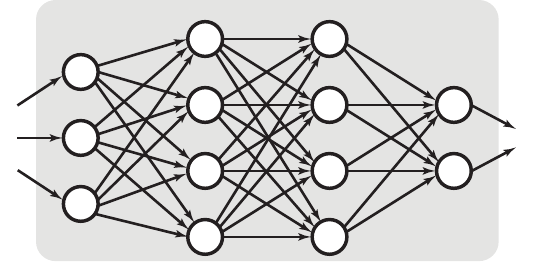}
		\put(7,0){\footnotesize Input layer}
		\put(30,0){\footnotesize Hidden layer 1}
		\put(54,0){\footnotesize Hidden layer 2}
		\put(79,0){\footnotesize Output layer}
		\put(18,46){$\hat{f}_1 ( \vect{x}_{0} ; \vect{\theta}_1  )$}
		\put(43,48){$\hat{f}_2 ( \vect{x}_{1} ; \vect{\theta}_2  )$}
		\put(68,45){$\hat{f}_3 ( \vect{x}_{2} ; \vect{\theta}_3  )$}
		\put(100,26){$\vect{y}$}
		\put(-3,26){$\vect{x}_0$}
\end{overpic} 
                \caption{A fully-connected feed-forward network with four layers ($L=3$) that fits into the box in (a).} 
        \end{subfigure} 
        \caption{The gray-box input-output model in (a) is characterized by $\hat{f}$ and a parameter vector $\vect{\theta}$. It is called an artificial neural network if $\hat{f}$ has a particular structure, such as the one illustrated in (b).}
        \label{fig:neural_network}  
\end{figure}
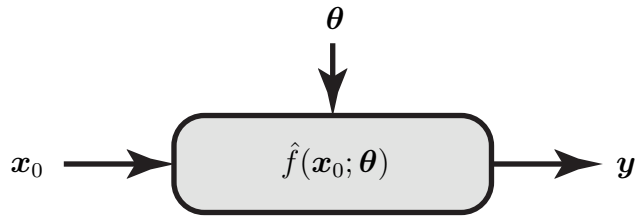
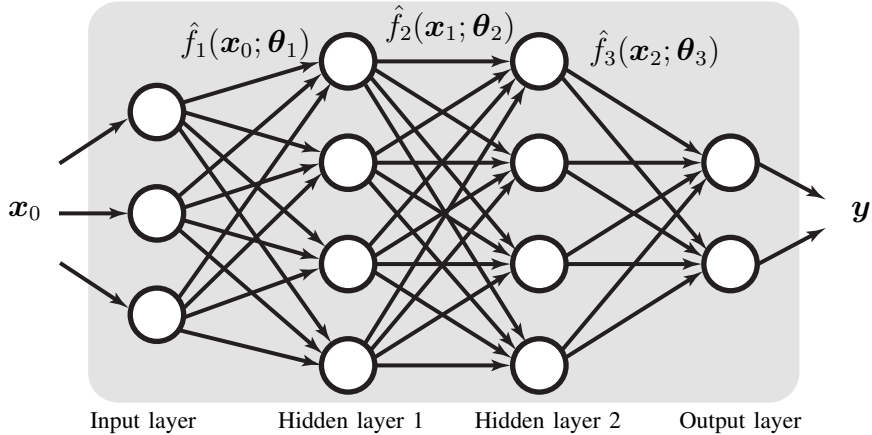


\section{Problem Statement and Solution}

{We begin by briefly describing what artificial neural networks are and formulating the problem of using them as function approximators.}

Consider a system that takes an $n_0$-length input vector $\vect{x}_0 \in \mathbb{R}^{n_0}$ and produces a $k$-length output vector $\vect{y} \in \mathbb{R}^{k}$, as illustrated in Fig.~\ref{fig:neural_network}(a).
The output is determined by the input via a deterministic function $\hat{f}$: 
\begin{equation}
\vect{y} = \hat{f}(\vect{x}_0;\vect{\theta}).
\end{equation}
The function is fixed but is characterized by an $m$-dimensional parameter vector $\vect{\theta} \in \mathbb{R}^{m}$. Many different input-output relations can be modeled in this way by changing the parameter vector $\vect{\theta}$, but they all share an underlying structure determined by the initial choice of $\hat{f}$. This is called a \emph{gray-box model}.

When the function $\hat{f}$ is selected to resemble the biological neural networks in human brains, the gray box is called an \emph{artificial neural network}. The input vector $\vect{x}_0$ is then viewed as the values in $n_0$ neurons from which the function $\hat{f}$ produces the values of $\vect{y}$ in $k$ other neurons.
 There are many different examples of this. The classical one is a \emph{fully-connected feed-forward network}, which is illustrated in Fig.~\ref{fig:neural_network}(b). In this case, $\hat{f}$ is a composition of $L$ functions, $\hat{f}_1,\ldots,\hat{f}_L$, which describes transitions between neurons in an input layer to neurons in an output layer via $L-1$ intermediate ``hidden'' layers. $L$ characterizes how \emph{deep} the network is. The function $\hat{f}_l$ is determined by the parameters $\vect{\theta}_l = \{ \vect{W}_l, \vect{b}_l \}$ and modeled as
\begin{equation} 
\hat{f}_l ( \vect{x}_{l-1} ; \vect{\theta}_l  ) = \sigma_l ( \vect{W}_l \vect{x}_{l-1} + \vect{b}_l ),
\end{equation}
where $ \vect{W}_l \in \mathbb{R}^{n_l \times n_{l-1}}$ is called a weight matrix, $\vect{b}_l \in \mathbb{R}^{n_l}$ is called a bias vector, and $\sigma_l : \mathbb{R}^{n_l} \to \mathbb{R}^{n_l}$ is an element-wise non-linear function that is called an activation function. With inspiration from the structure of the human brain, the function $\hat{f}_l$ can be interpreted as taking the values $\vect{x}_{l-1}$ in the $n_{l-1}$ neurons of layer $l-1$, mixing the values together according to the affine transition relation $\vect{W}_l \vect{x}_{l-1} + \vect{b}_l $, and finally applying the activation function $\sigma_l$ to the determine values of the $n_l$ neurons of layer $l$.

If there are four layers as in Fig.~\ref{fig:neural_network}(b), then $L=3$ and the complete input-output relation is
\begin{equation}
\vect{y} = \hat{f}_3 \left( \hat{f}_2 \left(  \hat{f}_1 \left( \vect{x}_0  ; \vect{\theta}_1 \right); \vect{\theta}_2 \right) ; \vect{\theta}_3 \right).
\end{equation}
Hence, the composite function $\hat{f}$ is determined by the parameter vector $\vect{\theta}$ containing the $\sum_{l=1}^{L} n_{l} (n_{l-1} +1)$
parameter values from $\vect{\theta}_1,\vect{\theta}_2,\vect{\theta}_3$ (i.e., the weights and biases from all layers).

\subsection{Problem Statement}

Artificial neural networks are generally used to approximate other functions, by selecting the parameter vector $\vect{\theta}$ to somehow minimize the approximation error. In particular, the category of fully-connected feed-forward networks is capable of approximating any continuous function arbitrarily well by utilizing a (possibly) large but finite number of parameters (and neurons)  \cite{Cybenko1989}.
This important result can be viewed as a generalization of Taylor polynomial approximations to functions with vector inputs and vector outputs. Two other categories are convolutional neural networks and recurrent neural networks \cite{Goodfellow-et-al-2016}. Each category is believed to be better at approximating certain types of functions, in the sense of requiring fewer parameters to achieve a certain approximation error and/or it being easier to find appropriate parameter values in practice. 

Selecting the right category of neural network is important but beyond the scope of this lecture note.
{Instead, our problem statement is: what are the important use cases where the function approximation capability can be utilized in the physical layer of communication systems, to achieve large improvements compared to conventional techniques?}



\subsection{Solution}

{To identify practically important use cases, we first need to understand how the function approximation is carried out.}
The parameter vector of an artificial neural network can be tuned/trained to approximate a (possibly unknown) function that we call $f$; that is, $\hat{f}$ should be trained to become a good estimate of $f$. This is preferably done by \emph{supervised learning} using a set of $T$ training examples consisting of input vectors $\vect{x}_t^{\train}$ and the corresponding output vectors $\vect{y}_t^{\train} = f(\vect{x}_t^{\train})$ that we want the neural network to reproduce, for $t=1,\ldots,T$.
Let us represent these training examples as the columns of two matrices:
\begin{align}
\vect{X}^{\train} = \begin{bmatrix} \vect{x}_1^{\train} &  \ldots & \vect{x}_T^{\train} \end{bmatrix}, \quad \quad
\vect{Y}^{\train} = \begin{bmatrix} \vect{y}_1^{\train}  & \ldots & \vect{y}_T^{\train}\end{bmatrix}.
\end{align}
The inputs should ideally be selected independently at random from the distribution of inputs that appears when using $f$ in reality. The training basically consists of finding the parameter $\vect{\theta}^*$ that minimizes a \emph{loss function} $\ell$ that measures the approximation mismatch:
\begin{equation} \label{eq:training-network}
\vect{\theta}^* = \argmin{\vect{\theta}} \,\, \ell \left( \vect{\theta}, \vect{X}^{\train},  \vect{Y}^{\train} \right).
\end{equation}
For example, the loss can be measured in the mean-squared sense as
\begin{equation}
\ell \left( \vect{\theta}, \vect{X}^{\train},  \vect{Y}^{\train} \right) = \frac{1}{T} \sum_{t=1}^{T} \left\| \vect{y}_t^{\train} - \hat{f}(\vect{x}_t^{\train};\vect{\theta})  \right\|^2.
\end{equation}
The goal is that the trained neural network $\hat{f}(\vect{x}_0;\vect{\theta}^*)$ will provide approximately the right outputs not only for the training examples, but for any input signal $\vect{x}_0$ generated in the same way. This desired property is called \emph{generalization}. 
Intuitively, if the unknown function $f$ is continuous and has limited variability, we should be able to approximate it well from a large training set. We can once again make a parallel to polynomial approximations; any scalar polynomial of order $T-1$ is uniquely determined by $T$ samples (training examples) of the inputs and outputs. If the polynomial order is unknown, or if the function is only approximately polynomial, we need a larger number of samples to ensure a good approximation.

Since the training in \eqref{eq:training-network} is a complicated non-convex optimization problem, huge efforts have been dedicated to finding computationally and performance-wise acceptable suboptimal solutions. Moreover, the generalization to unseen inputs can be improved by various regularizations, hyper-parameter choices, and network designs \cite{Goodfellow-et-al-2016}. These choices affect the model complexity. A simple model cannot capture complex dependencies. A too complex model explains the training data only (this is called overfitting). A correct complexity trade-off gives good generalization and is typically found using cross-validation. However, such empirical craftsmanship is not the focus of this lecture note, but we conclude:

\begin{enumerate}

\item Artificial neural networks can approximate any continuous function.

\item The supervised training requires a large training set with inputs/outputs to achieve a low approximation error.

\end{enumerate}


{There are many functionalities in  communication systems that can be described by a mathematical function $f$ and, thus, can be approximated by a neural network. To identify the promising use cases, we will first explain the basic methodology and its weaknesses by giving a concrete example.}

\subsubsection{A Deep-Learning Solution to Signal Detection}

The physical layer of a communication system determines how an information-bearing signal is sent from the transmitter to the receiver over a physical channel. A critical task is the signal detection, where the receiver tries to identify what information was sent. To describe some key properties of deep learning, we will exemplify how it can be used for signal detection.

\begin{figure} 
        \centering
        \begin{subfigure}[t]{\columnwidth} \centering \vspace{-3mm}
	\begin{overpic}[width=.45\columnwidth,tics=10]{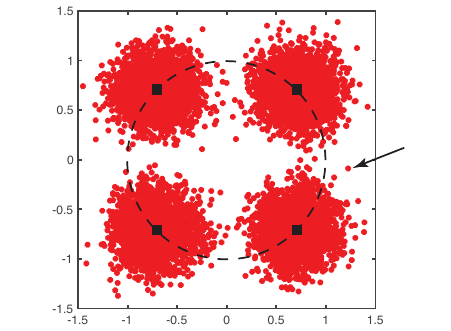}
		\put(31,55.5){\footnotesize$01$}
		\put(66,55.5){\footnotesize$11$}
		\put(66,18){\footnotesize$10$}
		\put(31,18){\footnotesize$00$}
		\put(91.5,39){A received signal $\vect{r}$}
\end{overpic} \vspace{-3mm}
                \caption{Quadrature phase-shift keying for information encoding and the corresponding received signals.} 
        \end{subfigure}
        \begin{subfigure}[t]{0.45\columnwidth} \centering 
	\begin{overpic}[width=\columnwidth,tics=10]{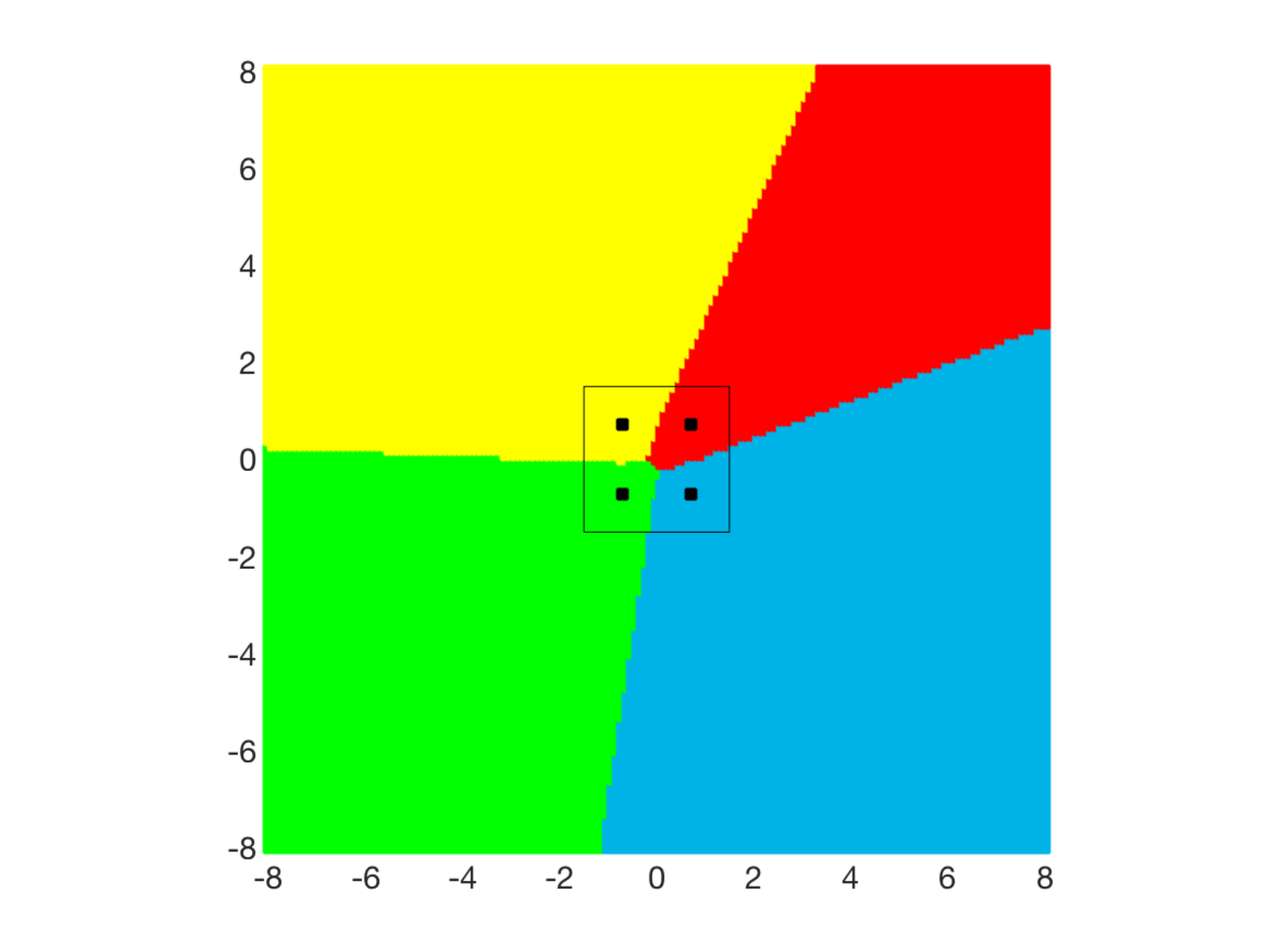}
		\put(23,56){\footnotesize Detection: $01$}
		\put(60,56){\footnotesize Detection: $11$}
		\put(60,18){\footnotesize Detection: $10$}
		\put(23,18){\footnotesize Detection: $00$}
\end{overpic} \vspace{-12mm}
                \caption{Detection regions with a trained neural network.} 
        \end{subfigure}
        \begin{subfigure}[t]{0.45\columnwidth} \centering
	\begin{overpic}[width=\columnwidth,tics=10]{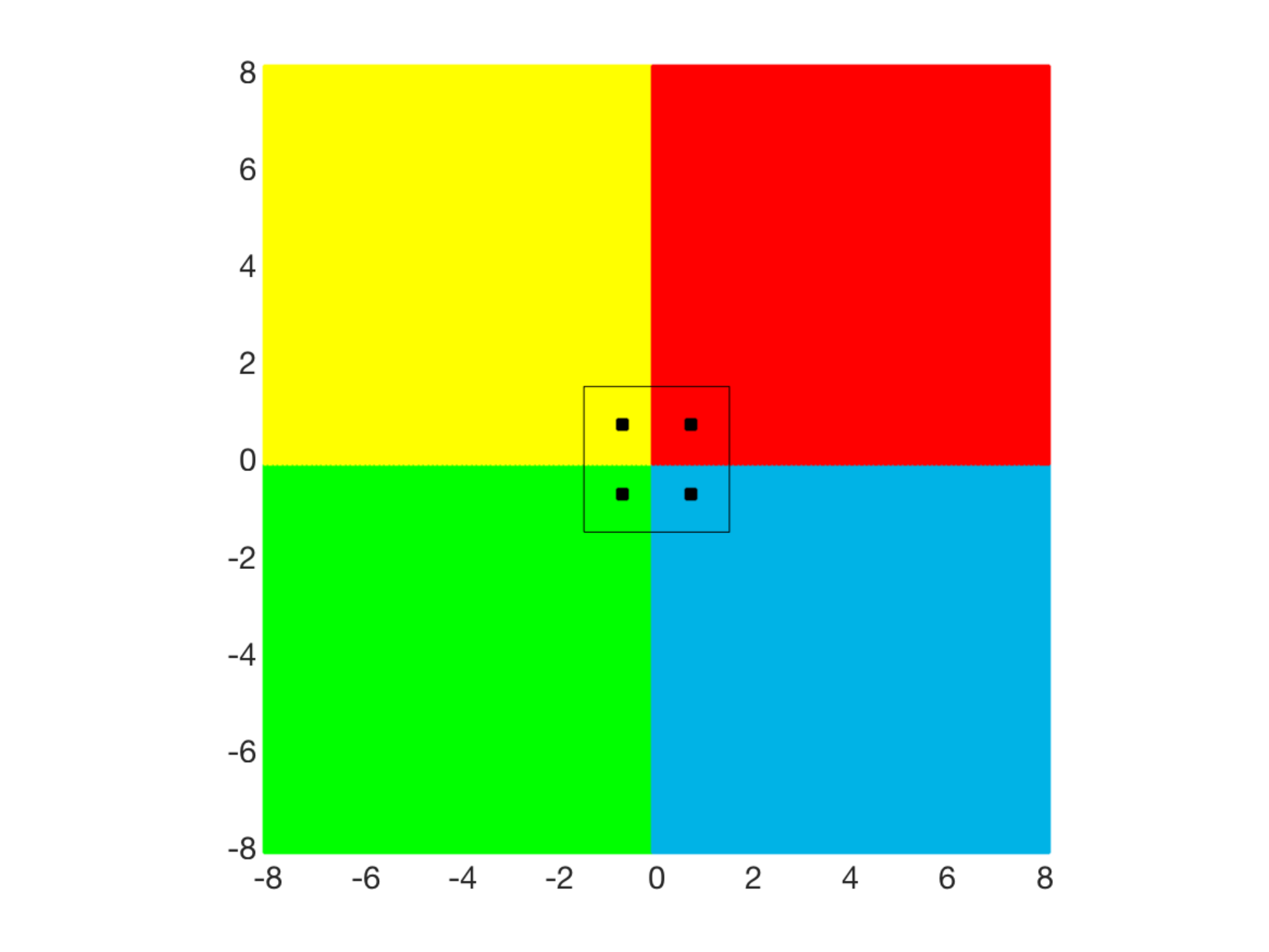}
		\put(23,56){\footnotesize Detection: $01$}
		\put(60,56){\footnotesize Detection: $11$}
		\put(60,18){\footnotesize Detection: $10$}
		\put(23,18){\footnotesize Detection: $00$}
\end{overpic}  \vspace{-12mm}
                \caption{Optimal detection regions using detection theory.} 
        \end{subfigure} 
        \caption{We send QPSK signals over an AWGN channel, as shown in (a), and try to detect the signals at the receiver. The detection regions produced by a trained neural network is shown in (b) and the optimal regions obtained from detection theory are shown in (c).}
        \label{fig:detection_example}   \vspace{-2mm}
\end{figure}
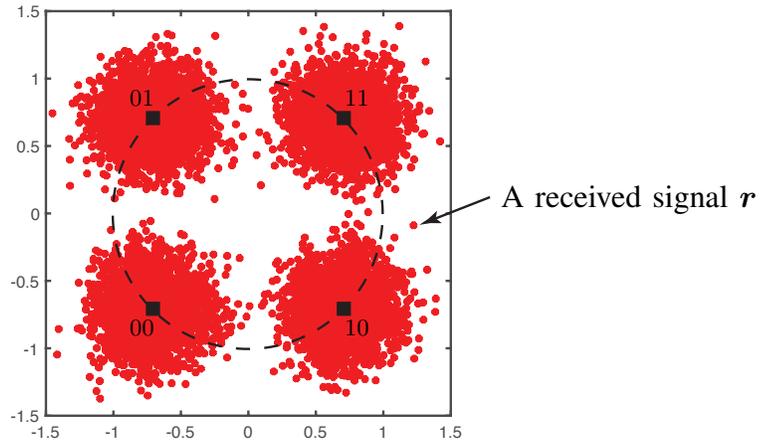
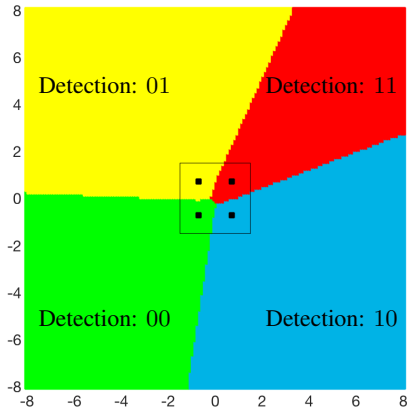
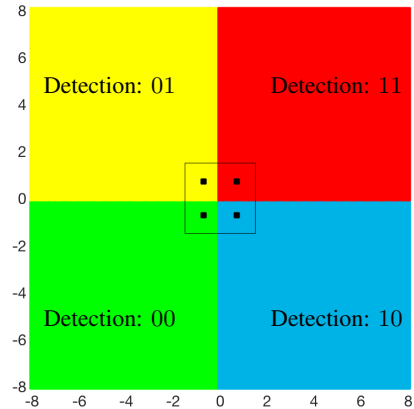

We consider a classical additive white Gaussian noise (AWGN) channel, where a two-dimensional signal vector $\vect{s} \in \mathbb{R}^2$ is sent. The received signal $\vect{r} \in \mathbb{R}^2$ is given by
\begin{equation}
\vect{r} = \vect{s} + \vect{n},
\end{equation}
where $\vect{n} \sim \mathcal{N}(\vect{0},\sigma^2\vect{I})$ is an independent Gaussian noise vector where the entries have variance $\sigma^2$. 
We assume two bits of information are encoded into $\vect{s}$ using a quadrature phase-shift keying (QPSK) constellation. Hence, there are four possible signal points that are equally spaced on the unit circle:
\begin{equation} \label{eq:QPSK}
\vect{s} \in \left\{  \begin{bmatrix} 1/\sqrt{2} \\ 1/\sqrt{2} \end{bmatrix}, \begin{bmatrix} 1/\sqrt{2} \\ -1/\sqrt{2} \end{bmatrix} , \begin{bmatrix} -1/\sqrt{2} \\ 1/\sqrt{2} \end{bmatrix} , \begin{bmatrix} -1/\sqrt{2} \\ -1/\sqrt{2} \end{bmatrix}   \right\}.
\end{equation}
The mapping between information bits and signals is illustrated in Fig.~\ref{fig:detection_example}(a). Due to the additive noise, the received signal $\vect{r}$ can take any value, but the Gaussian distribution makes values close to one of the signal points in \eqref{eq:QPSK} more likely than values far away. This can be seen from the red dots in Fig.~\ref{fig:detection_example}(a), which represent $\vect{r}$ for 10,000 noise realizations with $\sigma^2 = 0.2$ that are added to each signal point.

Based on the received signal $\vect{r}$, the receiver needs to guess (detect) what signal $\vect{s}$ was sent. We have trained a neural network for this task, by taking the received signal $\vect{x}_0=\vect{r}$ as input and letting the output $\vect{y}$ be a four-dimensional vector that is one for the detected signal and has zeroes elsewhere.
We used the 40,000 red dots in Fig.~\ref{fig:detection_example}(a), and the signals $\vect{s}$ that generated these $\vect{r}$, to train a fully-connected neural network using standard training methods.
We then applied the neural network to a wide range of possible received signals to illustrate how it is making its detection. The colored areas in Fig.~\ref{fig:detection_example}(b) show in which regions the received signals are mapped to the respective information signals. The regions are separated by lines, which is expected since each layer performs linear algebra operations; in particular, each activation function determines if the input is below/above a line that has been selected by training. Note that we have ``zoomed out'' and the range of values that was shown in Fig.~\ref{fig:detection_example}(a) is indicated by the black square.

The colored detection regions produced by the neural network have peculiar asymmetric shapes, which are not optimal. In fact, the optimal detection regions for AWGN channels are well known \cite[Ch.~6]{madhow_2014}: the received signal should be mapped to the closest signal point in terms of Euclidean distance. The optimal detection regions  are shown in Fig.~\ref{fig:detection_example}(c). The regions are quite similar within the black square, but greatly deviates further away. Several important observations can be made from this example:

\begin{enumerate}

\item If there is a known optimal algorithm, a trained neural network cannot outperform it. The detection error probability is, however, almost the same in this particular example since most received signals appear within the black square where the neural network has a decent behavior.

\item There are two reasons why the detection regions in Fig.~\ref{fig:detection_example}(b) are wrongly shaped.
Firstly, the shape inside the black square (around the signal points) is wrong due to overfitting; the training examples in Fig.~\ref{fig:detection_example}(a) can be approximately separated by many piecewise linear boundaries, including the ones shown in Fig.~\ref{fig:detection_example}(b).
Secondly, since all training examples are inside the  square, the behavior outside the square is somewhat random; the neural network has learned to interpolate between training examples but not to extrapolate outside the square. This is a practical issue since received signals far outside the square occasionally appear due to the long-tailed Gaussian distribution.
 This is a general phenomenon; neural networks are good at handling typical inputs but may generalize poorly to atypical inputs.

\item We could have used prior domain knowledge (from digital communications) to preprocess the input signals. In this example, the neural network had to rediscover where the constellation points are, how the noise is distributed, and how to make the right detection.
If we would instead compute the Euclidean distance between the received signal and each of the four signal constellation points, we could use that as input to a neural network. This will give more accurate and reliable results since we have utilized our domain knowledge to ensure that the neural network has fewer characteristics to learn. However, it still cannot beat the optimal detection.
\end{enumerate}

\subsubsection{Is There a Role of Deep Learning in Communications?}

Since signal detection in AWGN channels is easy to perform optimally, it makes little sense to utilize artificial neural networks for that purpose. There are many similar tasks in communications where deep learning cannot make any meaningful improvements.
For example, the fundamental performance limits were derived by Shannon \cite{Shannon1948a} and we can operate close to those limits using modern channel codes. Moreover, it is known how to perform optimal channel estimation, multi-user multiple-input multiple-output (MIMO) processing, and transmit power allocation in many wireless communication scenarios \cite{massivemimobook}. The fact that the information signals are man-made gives us strong prior information that makes it easier to devise effective man-made algorithms than in many other fields, where the signals are created by nature.

There are nevertheless some important roles that deep learning can play in communications. Firstly, there are many problems where a known algorithm finds the optimum but has prohibitively high complexity for real-time implementation. Secondly, there are cases where the standard system models are inadequate or incomplete. It is sufficient to replace the noise distribution in the previous example with an unknown one to find a case where learning can help.
We will elaborate on these two applications in the remainder of this lecture note. 
But before that, we stress that errors are unavoidable in the physical layer of communication systems and are conventionally dealt with using retransmissions. This built-in fault tolerance is positive when it comes to the utilization of deep learning. It gives robustness to the strange behaviors that occasionally occur when an atypical signal is fed into a neural network that has been trained to work well for typical input signals.
However, adversaries can also exploit atypical signals to perform jamming more efficiently \cite{Sadeghi2019a}.


\section{Application 1: Algorithmic Approximation}


The first important application of deep learning in communications is to approximate a known but computationally complicated algorithm. There are many examples of iterative algorithms that asymptotically find a global (or local) optimum to an optimization problem, but require very many iterations for convergence and/or complicated operations in each iteration \cite{Sun2018a}. Such algorithms might not be practically useful in communication systems where latency constraints require execution times below a millisecond.

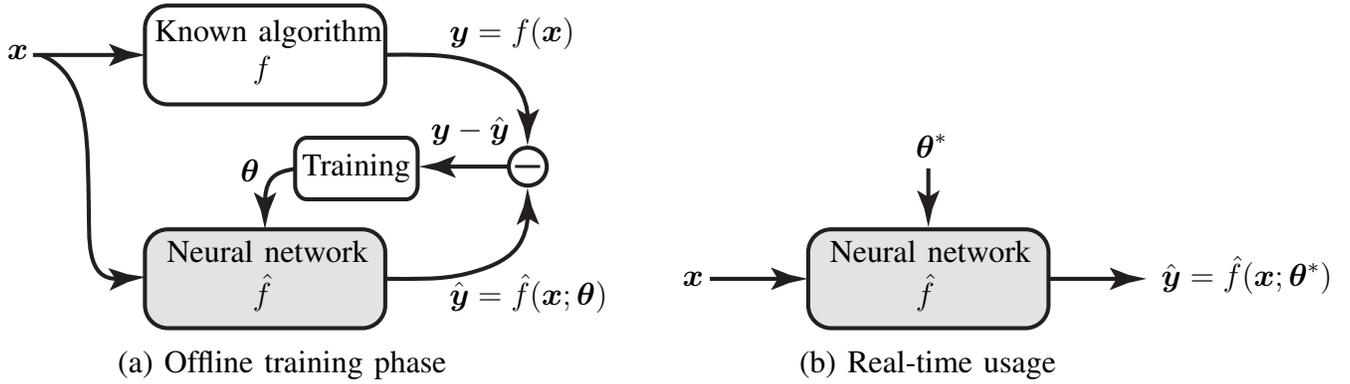
\begin{figure}  \vspace{-7mm}
	\begin{overpic}[width=0.9\columnwidth,tics=10]{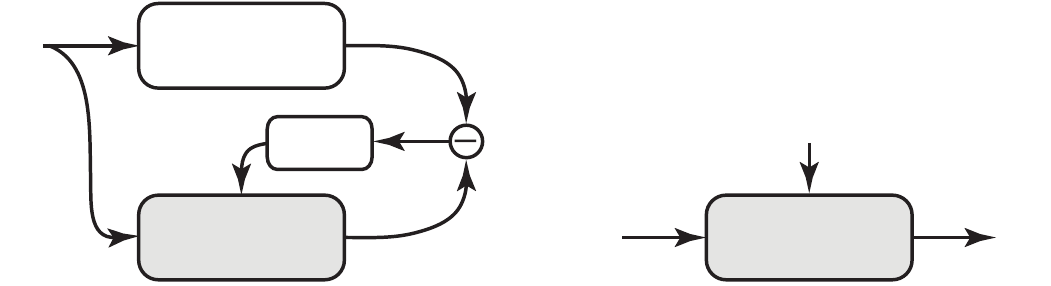}
		\put(14,24){Known algorithm}
		\put(22,21){$f$}
		\put(26,13){Training}
		\put(76,14.5){$\vect{\theta}^*$}
		\put(21,12.5){$\vect{\theta}$}
		\put(15,6){Neural network}
		\put(69,6){Neural network}
		\put(22,2.5){$\hat{f}$}
		\put(76,2.5){$\hat{f}$}
		\put(36.5,16){$\vect{y}-\hat{\vect{y}} $}
		\put(86,44.5){$\vect{y}=f(\vect{x}) $}
		\put(38,24){$\vect{y}=f(\vect{x})$}
		\put(38,2.5){$\hat{\vect{y}}=\hat{f}(\vect{x};\vect{\theta})$}
		\put(96,4){$\hat{\vect{y}}=\hat{f}(\vect{x};\vect{\theta}^*)$}
		\put(2,22.5){$\vect{x}$}
		\put(57,4){$\vect{x}$}
		\put(11,-3){(a) Offline training phase}
		\put(66.5,-3){(b) Real-time usage}
\end{overpic} 
\vspace{5mm}
        \caption{A known algorithm $f$ can be approximated by training a neural network $\hat{f}$ to make $f(\vect{x}) \approx \hat{f}(\vect{x};\vect{\theta}^*)$ for all possible inputs, as shown in (a). The training procedure will iteratively update $\vect{\theta}$ to gradually reduce the approximation errors until it converges to some $\vect{\theta}^*$. 
        If the neural network is designed to have sufficiently low complexity, then the trained neural network in (b) can be used in real-time applications.} \vspace{-3mm}
        \label{fig:application1}  
\end{figure}

The general procedure for training a neural network for algorithmic approximation is illustrated in Fig.~\ref{fig:application1}. Suppose we have a known algorithm, represented by the function $\vect{y}=f(\vect{x})$, which cannot be implemented in real time. To address this problem using deep learning, we can first create a training set containing a large number $T$ of input signals $\vect{x}_t^{\train}$, for $t=1,\ldots,T$. We then run the algorithm $T$ times to compute the outputs 
\begin{equation} \label{eq:training-set-generation}
\vect{y}_t^{\train} = f(\vect{x}_t^{\train}).
\end{equation}
After having generated the training set, we can train an artificial neural network to provide approximately the same outputs for these inputs. More precisely, we should find an optimized parameter vector $\vect{\theta}^*$ in accordance to \eqref{eq:training-network}. If the training is performed well, the neural network will  generalize well (i.e., provide good outputs) to previously unseen input signals that were generated in the same way as the inputs used for training. Simply speaking, this means that $f(\vect{x}) \approx \hat{f}(\vect{x};\vect{\theta}^*)$ for all inputs $\vect{x}$ of practical interest.

There are many optimization problems to be solved in communication systems. For example, at the transmitter, power allocation between concurrent transmissions is important to limit interference \cite{massivemimobook,Sun2018a}. At the receiver, non-linear signal detection problems must be solved to deal with interference in MIMO systems \cite{Samuel2017a}. Some of these problems are convex and can be solved by off-the-shelf optimization software. Other problems are non-convex but there exist iterative algorithms that converge to local or global optima. In both cases, the computational complexity is often prohibitive for real-time applications, where similar optimization problems with different input data are solved repeatedly. A neural network can then be trained to learn approximately how the solution depends on the input data. This approximate input-output map can be evaluated with substantially lower computational cost, as exemplified in \cite{Sun2018a,Samuel2017a}. Domain knowledge can be utilized to pre-process the input data, to focus the learning on the  problem that the algorithm is solving and not on rediscovering known properties (e.g., that the desired signal lies in a certain subspace).

There are two main approaches. One can learn the input-output mapping based on training data, as described above, while ignoring how it was produced. Alternatively, the shape of the neural network can be selected so that each layer mimics one iteration of a known algorithm that converges asymptotically to an optimum. This is called  \emph{deep unfolding} and exploits that many first-order iterative optimization methods have the same structure as a (recurrent) neural network \cite{Hershey2014a}. The parameters of the neural network are then trained to give a nearly optimum solution after a predefined number of iterations, thereby speeding up the convergence. In \cite{Samuel2017a}, the authors ``unfold'' a gradient-descent-like algorithm for MIMO detection to create a neural network where each layer performs similar operations but with optimized parameters. When using this approach, \eqref{eq:training-set-generation} needs not to be determined in advance, which simplifies the training.
 
The practical benefit of this application is the complexity reduction it can provide; the neural network will essentially learn how to make algorithmic shortcuts to strike a good balance between accuracy and computational complexity. Another important benefit is related to hardware implementation. To solve a practical problem with real-time constraints, we conventionally would first need to design an algorithm and then develop a dedicated circuit based on it, which can be very time-consuming. With the help of deep learning, we can instead predesign a general-purpose circuit that implements a neural network of a given maximum size (i.e., number of layers and neurons) with a predetermined run time. We can then train a neural network to perform the algorithmic task we need and, finally, load the corresponding trained parameters (i.e., weights and biases) onto the circuit. This new approach to hardware implementation can greatly reduce the time from that the algorithmic design begins to a product can hit the market.

A main issue with this application is the highly computationally demanding generation of desired outputs: the more complex the algorithm $f$ is, the longer time it takes to compute $f(\vect{x}_t^{\train})$ for $t=1,\ldots,T$. We are basically moving the  complexity issue from the algorithmic run time to the design process. There is a practical limit to which algorithms that we can approximate in this way. If it takes 1 hour to generate one training example, it will take 11.4 years or extreme parallelism to generate 100,000 examples.


\section{Application 2: Inversion of an Unknown Function}

The second important application is to invert an unknown function. In particular, non-linear distortion can occur between the transmitter and receiver. Three prominent examples are finite-resolution quantization in the receiver hardware, non-linear amplifiers in the transmitter hardware \cite{Demir2019a}, and non-linear fiber-optical channels \cite{Ellis2010a}. While quantizers typically are designed with known properties, the latter two examples can be represented by an unknown function $g$ that takes a signal $\vect{y}$ as input and produces a distorted output $\vect{x}=g(\vect{y})$. The conventional way to undo the distortion is to identify an appropriate parameterized model of the function, then estimate the parameters from measurements, and finally create an inverse function based on the estimates. This three-step approach is suboptimal and prone to error-propagation. An alternative is to train a neural network to directly invert the function, without explicit modeling or parameter estimation. Whenever only suboptimal conventional algorithms exist, a learned algorithm can theoretically provide better performance and robustness, but only if the training is carried out successfully.

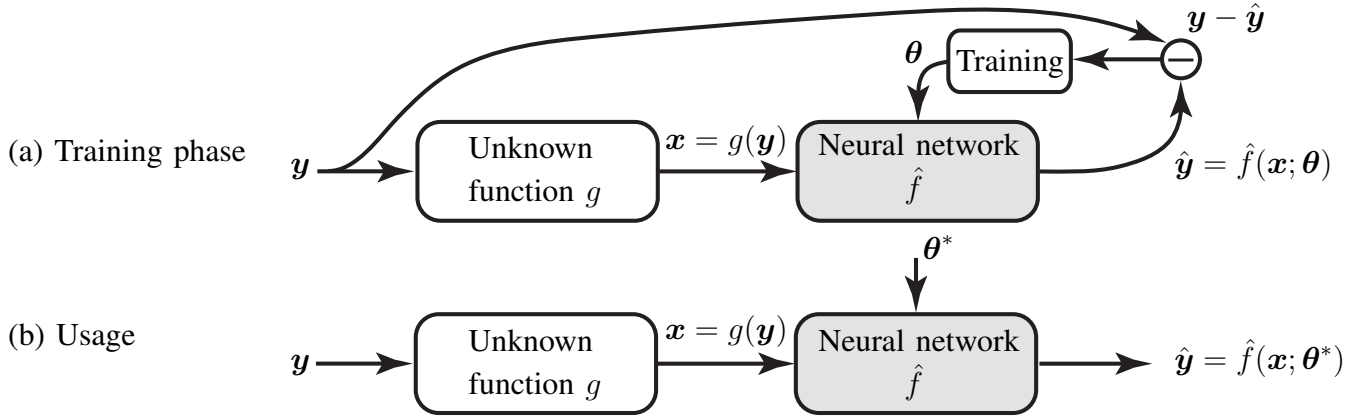
\begin{figure}  \vspace{-4mm}
	\begin{overpic}[width=0.9\columnwidth,tics=10]{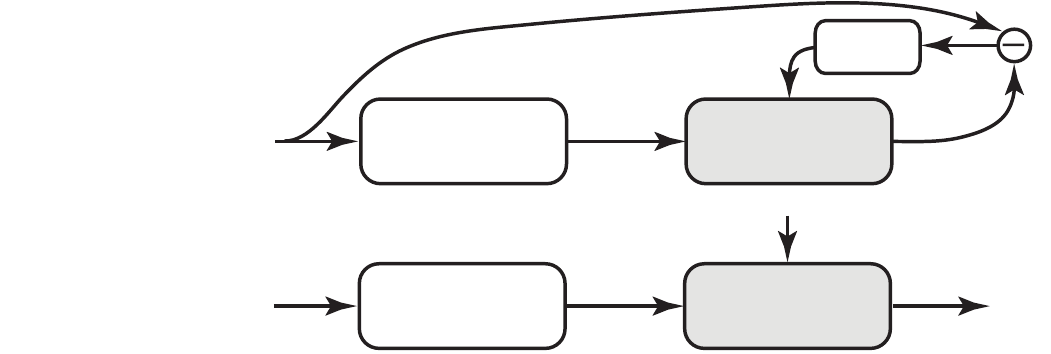}
		\put(78.2,28.3){Training}
		\put(74,29){$\vect{\theta}$}
		\put(75.5,13){$\vect{\theta}^*$}
		\put(38.5,21.2){Unknown}
		\put(38.5,5.5){Unknown}
		\put(38.5,17.7){function $g$}
		\put(38.5,2){function $g$}
		\put(67,21.5){Neural network}
		\put(67,5.5){Neural network}
		\put(74,18){$\hat{f}$}
		\put(74,2){$\hat{f}$}
		\put(97,32){$\vect{y}-\hat{\vect{y}} $}
		\put(96,20){$\hat{\vect{y}}=\hat{f}(\vect{x};\vect{\theta})$}
		\put(96,4){$\hat{\vect{y}}=\hat{f}(\vect{x};\vect{\theta}^*)$}
		\put(54.5,22){$\vect{x}=g(\vect{y})$}
		\put(54.5,6.5){$\vect{x}=g(\vect{y})$}
		\put(24,20){$\vect{y}$}
		\put(24,4){$\vect{y}$}
		\put(1,21){(a) Training phase}
		\put(1,6){(b) Usage}
\end{overpic} 
        \caption{An unknown function $g$ with input $\vect{y}$ is inverted using a neural network $\hat{f}$ by training it to achieve $\hat{f}(g(\vect{y});\vect{\theta}^*) \approx \vect{y}$, as shown in (a).
        The training procedure will iteratively update $\vect{\theta}$ to gradually reduce the approximation errors until it converges to some $\vect{\theta}^*$. 
        The trained neural network in (b) can be used to counteract the unknown function, without having to explicitly model it and estimate model parameters.}
        \label{fig:application2}  
\end{figure}

The general procedure for training a neural network for function inversion is illustrated in Fig.~\ref{fig:application2}. We need to generate a large number $T$ of possible communication signals $\vect{y}_t^{\train}$ and send them through the unknown function to measure 
\begin{equation}
\vect{x}_t^{\train} = g(\vect{y}_t^{\train}) \quad \textrm{for } t=1,\ldots,T.
\end{equation}
It is then $\vect{x}_t^{\train}$ that is used as input to the neural network, while $\vect{y}_t^{\train}$ is the desired output.

Different from Application 1, the creation of a training set can be very computationally efficient in Application 2 because the outputs are man-made. It is typically created to be statistically equivalent to the signals observed at run time, but one can also create a biased training set to emphasize typical or atypical examples.
Online learning when operating the communication system is possible by occasionally sending predefined reference signals to generate new training data. This is useful when the function $g$ is time-varying (e.g., due to temperature variations in the hardware).
The key to successful utilization of deep learning is to identify tasks in communication systems that currently lack an optimal solution---there is then an opportunity to beat the state-of-the-art. For example, a common way to deal with non-linear communication hardware is to apply the Bussgang decomposition \cite{Bussgang1952a} to write the output of the non-linear function $g$ as $g(\vect{y}) = \vect{D} \vect{y} + \boldsymbol{n}$, where $\vect{D} $ is a deterministic matrix and $\boldsymbol{n}$ is distortion noise that is uncorrelated with $\vect{y}$ but statistically dependent. By pretending as if $\boldsymbol{n}$ is independent noise, one can often develop communication algorithms (e.g., for channel estimation or data detection) that partially mitigate distortion, but such algorithms are suboptimal since the distortion is in fact dependent on the input. As shown in \cite{Demir2019a}, one can  achieve substantially better performance by training neural networks instead.

\section{What We Have Learned}

Although many parts of communication systems can be solved optimally, there are important cases where deep learning can give large improvements. In particular, it can be used to reduce computational complexity of known algorithms or to deal with non-linear hardware or channels in an efficient way.

\section{Acknowledgment}

This work was supported by the Wallenberg AI, Autonomous Systems and Software Program (WASP) funded by the Knut and Alice Wallenberg Foundation.

\section{Authors}

\textbf{Emil Bj\"ornson} (emil.bjornson@liu.se) received the MSc degree in engineering mathematics from Lund University, Sweden, in 2007, and the PhD degree in telecommunications from the KTH Royal Institute of Technology, Sweden, in 2011. He is now an associate professor at Link\"oping University, Sweden.
He has authored the textbooks \emph{Optimal Resource Allocation in Coordinated Multi-Cell Systems} (2013) and \emph{Massive MIMO Networks: Spectral, Energy, and Hardware Efficiency} (2017). He received the 2018 IEEE Marconi Prize Paper Award in Wireless Communications, the 2019 EURASIP Early Career Award, the 2019 IEEE Communications Society Fred W. Ellersick Prize, and the 2019 IEEE Signal Processing Magazine Best Column Award.

\textbf{Pontus Giselsson} (pontus.giselsson@control.lth.se) is an Associate Professor at the Department of Automatic Control at Lund University, Sweden. His current research interests include mathematical optimization and its wide range of applications, e.g., in machine learning, control, signal processing, and wireless communication. He received an MSc degree from Lund University in 2006 and a PhD degree from Lund University in 2012. During 2013 and 2014, he held a postdoc position at Stanford University. In 2012, he received the Young Author Price at the ADCHEM IFAC Symposium, in 2014, he received the Young Author Price at the IFAC World Congress, and in 2015, he received the Ingvar Carlsson Award from the Swedish Foundation for Strategic Research.



\bibliographystyle{IEEEtran}
\bibliography{IEEEabrv,refs}

\end{document}